\documentclass[aps,tightenlines,floats,amsmath,amssymb,nofootinbib,pra,showpacs,twocolumn]{revtex4-1}
\usepackage{amsmath}
\usepackage{amsfonts}
\usepackage{amssymb}
\usepackage{mathrsfs}
\usepackage{graphicx}

\begin{document}

\title{Ultimate Capacity of a Linear Time-Invariant Bosonic Channel}

\author{Bhaskar Roy Bardhan}
\email{broyba1@mit.edu}
\author{Jeffrey H. Shapiro}
\address{Research Laboratory of Electronics, Massachusetts Institute of Technology, Cambridge, Massachusetts 02139, USA}
\date{\today }

\begin{abstract}

We determine the ultimate classical information capacity of a linear time-invariant bosonic channel with additive phase-insensitive Gaussian noise. This channel can model fiber-optic communication at power levels below the threshold for significant nonlinear effects. We provide a general continuous-time result that gives the ultimate capacity for such a channel operating in the quasimonochromatic regime under an average power constraint. This ultimate capacity is compared with corresponding results for heterodyne and homodyne detection over the same channel.

\end{abstract}


\maketitle

\section{Introduction}

In the 1940s, Shannon developed a mathematical theory that delineates the maximum rate
at which error-free communication is possible over a communication channel~\cite{Shannon}. 
His seminal work revolutionized the understanding of information processing and has played a crucial role in the emergence of the information era.  Information transmission, however, relies on physical encoding.  When that encoding is at optical frequencies---as in the Internet's fiber-optic backbone---quantum-mechanical noise sets the ultimate rate of reliable communication. It follows that determining the information-carrying capacity of noisy quantum communication channels is of considerable practical relevance. 

Bosonic channels provide quantum models for fiber-optic and free-space optical communication~\cite{weedbrook,braunstein}. The rate of reliable information transmission through such communication channels depends on the receiver configuration that is used to extract the encoded information. In particular, conventional optical communication receivers---viz., direct, homodyne, or heterodyne detection receivers---have different capacities because the quantum measurements they perform lead to different measurement statistics. Direct detection has superior photon efficiency (bits/photon)~\cite{Robinson}, so it is the preferred choice for photon-starved applications like the Lunar Laser Communication Demonstration~\cite{LLCD}. Homodyne and heterodyne detection, however, offer better spectral efficiency (bits/sec-Hz)~\cite{Li}, thus they are being pursued to maximize throughput in the Internet's fiber backbone.

The ultimate capacity of a bosonic channel is its Holevo capacity~\cite{Holevo}.  For an optical communication channel the Holevo capacity  will equal or exceed those of the conventional systems.  Until recently, the only bosonic channel whose Holevo capacity was known was the pure-loss channel, in which any attenuation between the transmitter and the receiver was accompanied by the minimum (vacuum-state) noise level needed to  preserve the Heisenberg uncertainty principle~\cite{HolevoPureLoss}. Now, however, with the proof of the minimum output-entropy conjecture~\cite{GPCH13,GHG13}, the Holevo capacities are known for all single-mode bosonic channels with phase-insensitive Gaussian noise. The results of these works were later applied  \cite{PMG14} to derive the capacities of bosonic communication channels that are affected by nonzero memory and Gaussian noise. The memory model from \cite{PMG14}---a cascade of identical discrete beam splitters or amplifiers---does not include the quantum version of the archetypal communication channel from classical information theory:  a general, continuous-time, linear-time invariant (LTI) filter followed by the addition of statistically-stationary Gaussian noise.  

In this paper, we remedy the preceding deficiency by deriving the Holevo capacity for an average-power constrained, quasimonochromatic bosonic channel comprised of a stable LTI filter---which, at a particular frequency, may be amplifying or attenuating---followed by additive phase-insensitive Gaussian noise arising from a thermal environment.   For comparison purposes, we also present the homodyne and heterodyne detection capacities for the same channel.  These results are then evaluated numerically for attenuator-amplifier and amplifier-attenuator unit cells, such as might be present in a fiber-optic system~\cite{HY92,Boyd11,Fearn}. 

The remainder of the paper is organized as follows. Section~\ref{capacity-intro} reviews prior results for the classical information capacity of bosonic channels.  Section~\ref{LTI-intro} introduces the LTI channel with thermal noise to be considered in what will follow. In Sec.~\ref{capacities} we derive the Holevo capacity for data transmission through that LTI channel and present its well-known homodyne and heterodyne capacities. In Sec.~\ref{capacities-cell} we consider a particular normalized shape for the LTI filter's frequency response and compare the capacities it implies for two amplifier-attenuator unit-cell configurations.  Section~\ref{conclusion} contains our concluding remarks.

\section{Classical information capacity of bosonic channels}
\label{capacity-intro}

A $K$-mode bosonic channel can be represented by $K$ quantized modes of the
 electromagnetic field in a tensor-product Hilbert space, $\mathcal{H}^{\otimes K}=\otimes_{k=1}^{K} \mathcal{H}_{k}$, 
with $K$ pairs of input and output photon annihilation operators $\{\,\hat{a}^{\rm in}_{k},\hat{a}^{\rm out}_{k} : 1\le k \le K\,\}$. For the single-mode attenuating and amplifying channels,
the channel input is an electromagnetic field mode with photon annihilation operator $\hat{a}_{\rm in}$, and the resulting channel output
is another field mode whose photon annihilation operator $\hat{a}_{\rm out}$ is given by the commutator-preserving transformations
\begin{equation}
\label{mode-transformation}
\hat{a}_{\rm out}=\left\{\begin{array}{ll}
\sqrt{\eta}\,\hat{a}_{\rm in}+\sqrt{1-\eta}\,\hat{a}_{\rm env},& \mbox{attenuating channel}\\[.05in]
\sqrt{\kappa}\,\hat{a}_{\rm in}+\sqrt{\kappa-1}\,\hat{a}_{\rm env}^\dagger,& \mbox{amplifying channel,}\end{array}\right.
\end{equation}
where $0 < \eta  \le 1$ is the attenuating channel's transmissivity, $1 < \kappa <\infty$ is the amplifying channel's gain, and $\hat{a}_{\rm env}$ is the photon annihilation operator corresponding to an environmental-noise mode.

The pure-loss channel is an attenuating channel that injects the 
minimum quantum noise required to preserve the Heisenberg uncertainty principle, i.e., the $\hat{a}_{\rm env}$ mode is in its vacuum state. The thermal-noise channel is an attenuating channel whose $\hat{a}_{\rm env}$ mode is in a thermal state, viz., an isotropic Gaussian mixture of coherent states with average photon number $N_{\rm env} > 0$:
\begin{equation}
\hat{\rho}_{\rm env}=\int {\rm d}^{2} \alpha\, \frac{\exp(-|\alpha|^{2}/N_{\rm env})}{\pi N_{\rm env}} \,
|\alpha\rangle\langle\alpha|.
\label{thermalstate}
\end{equation}
The amplifying channel's $\hat{a}_{\rm env}$ mode injects minimal quantum noise when it is in its vacuum state, but, more generally, it too could be in a thermal state given by (\ref{thermalstate}).

Shannon's noisy channel coding theorem showed that the classical capacity of a classical channel is the maximum mutual information between its input and output over all encoding and decoding strategies. However, the quantum nature of the single-mode attenuating and amplifying channels we have just described means that their classical information capacities must be found from the Holevo, Schumacher, Westmoreland (HSW) theorem \cite{HSW1,HSW2}, specifically by maximizing the Holevo information over both the transmitted quantum states and the receiver's quantum measurement.   Consider a set of symbols $\{x \}$ that is represented by a collection of 
input states $\{\hat{\rho}_{x} \}$, and assume that these states are selected according to some prior distribution $\{ p_{x} \}$. A single use of a quantum channel---such as those governed by (\ref{mode-transformation})---can, in general, be represented by a 
completely-positive-trace-preserving map, $\mathcal{M}$, and the single-use Holevo information $\chi(\mathcal{M})$ for this channel is given by
\begin{equation}
\chi(\mathcal{M})=S \!\left ( \sum_{x} p_{x} \hat{\rho}_{x} \right )-\sum_{x} p_{x} S \left ( \hat{\rho}_{x} \right ),
\end{equation}
where $S(\hat{\rho})$ is the von Neumann entropy of a state $\hat{\rho}$.  According to the HSW theorem,
the classical capacity of this channel is
\begin{equation}
C_{\rm HSW}(\mathcal{M}) =\sup_{n} \left ( \max_{\{p_{x}, \hat{\rho}_{x} \} } \left [ \chi \left (\mathcal{M}^{\otimes n} \right ) \right ]/n \right ).
\end{equation}
The maximization in the above formula is performed over all input ensembles $\{ p_{x}, \hat{\rho}_{x} \}$, and the
regularization step---the supremum over $n$ channel uses---is necessary because Holevo information need not be additive. 

For the single-mode pure-loss bosonic channel $\mathcal{M}_{\rm pl}$, whose transmitter is constrained to use at most $N_{S}$ photons
on average per channel use, the HSW capacity (in bits/use) was shown to be additive and given by~\cite{HolevoPureLoss}
\begin{equation}
C_{\rm HSW}(\mathcal{M}_{\rm pl}) = g(\eta N_{S}),
\end{equation}
where
\begin{equation}
\label{eq:g-function}
g(x) \equiv (x+1)\log_{2}(x+1)-x\log_{2}(x)
\end{equation}
is the von Neumann entropy of a bosonic thermal state with average photon number $x$. Moreover, the same work showed that this capacity was achievable with an isotropic Gaussian encoding over coherent states.   This capacity exceeds 
what is achievable with coherent-state encoding and homodyne or heterodyne detection over the pure-loss channel, namely,
\begin{align}
C_{\text{hom}}(\mathcal{M}_{\rm pl})&=\frac{1}{2} \log_{2} (1+4 \eta N_{S})\\[.05in]
C_{\text{het}}(\mathcal{M}_{\rm pl})&=\log _{2}(1+\eta N_{S}),
\end{align}
with $C_{\text{het}}(\mathcal{M}_{\rm pl})/C_{\rm HSW}(\mathcal{M}_{\rm pl})\rightarrow 1$ as $N_{S} \rightarrow \infty$.

The pure-loss channel's HSW capacity was found in 2004, but it was only last year---with the proof of the minimum output-entropy conjecture \cite{GHG13,GPCH13}---that the following HSW capacities for the single-mode thermal-noise and amplifying channels were obtained:  
\begin{align}
C_{\rm HSW}(\mathcal{M}_{\rm therm}) &= g(\eta N_S + (1-\eta)N_{\rm env}) \nonumber \\[.05in]
&- g((1-\eta)N_{\rm env})\\[.05in]
C_{\rm HSW}(\mathcal{M}_{\rm amp}) &= g(\kappa N_S + (\kappa-1)(N_{\rm env}+1)) \nonumber \\[.05in]
&- g((\kappa-1)(N_{\rm env}+1)).
\end{align}
Note that they too are additive and achieved by isotropic Gaussian encoding over coherent states \cite{GHG13}.  The homodyne and heterodyne capacities for coherent-state communication over the single-mode thermal-noise and amplifying channels will be used in Sec.~\ref{capacities}, when we address the LTI channel's capacity.  

The HSW capacities for the single-mode bosonic channels can be extended to multi-mode channels. For multiple-spatial-mode, wideband, pure-loss channels, the ultimate limits on the capacity were derived in \cite{ultimate-free-space}, where it was shown that the capacity-achieving encoding employed all spatial modes and all frequencies. The results from \cite{GHG13,GPCH13} allow a further extension to include arbitrary multi-mode combinations of  thermal-noise and amplifying channels from (\ref{mode-transformation}). Thus, building on the single-mode capacity results for quantum attenuators and amplifiers, \cite{PMG14} evaluated the capacity of a specific Gaussian thermal memory channel model by considering its singular-value decomposition. That paper's memory model, however, is rather limited in its scope.  Providing a more inclusive treatment of bosonic memory channels with additive Gaussian noise is therefore the goal of the present paper.

\section{Linear time-invariant bosonic channel}
\label{LTI-intro}
\begin{figure}
\centering
	\includegraphics[width=\columnwidth]{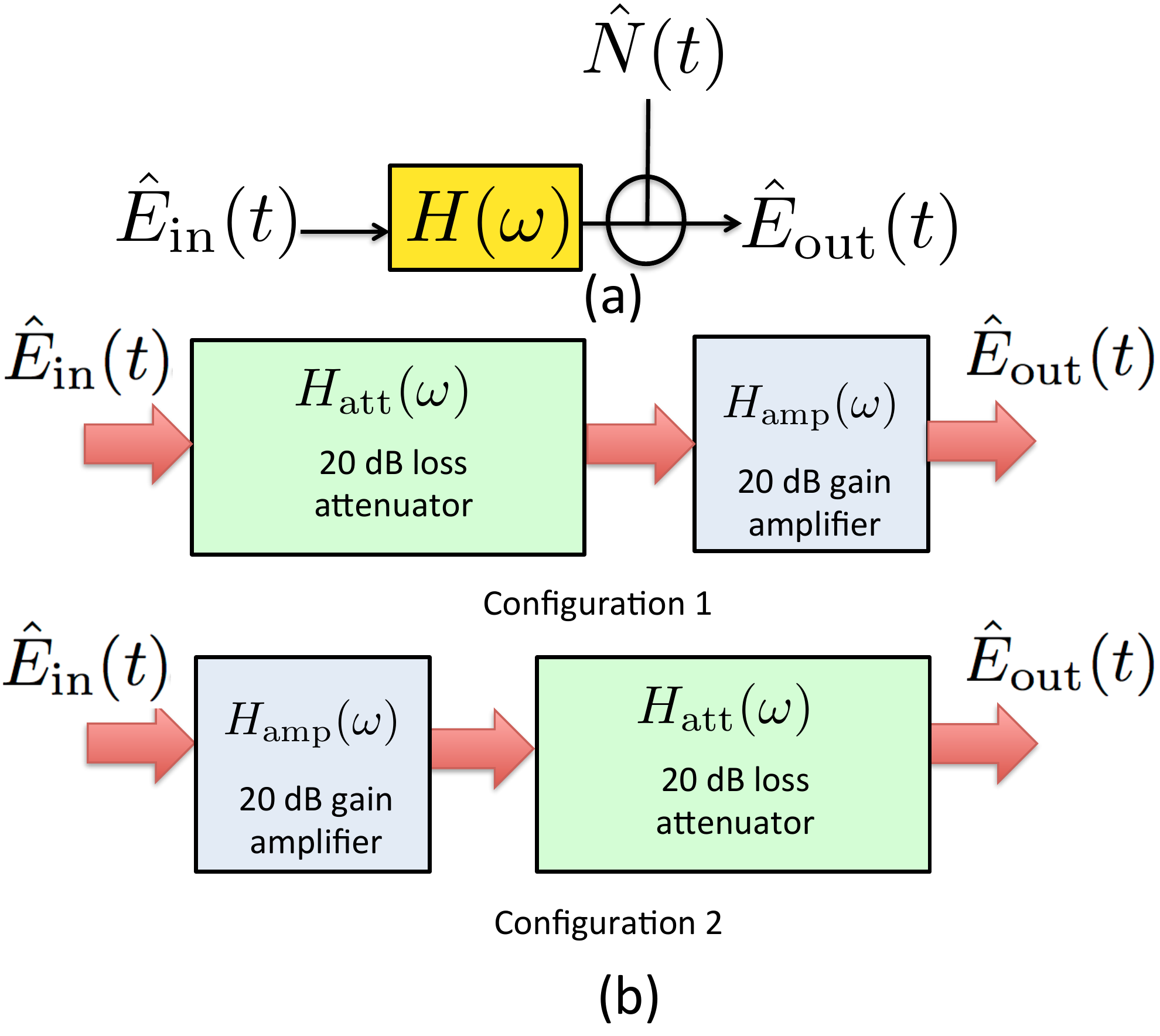}
	\caption{(color online). (a) Schematic for the transmission of baseband field operators through a channel comprised of an LTI filter---with frequency response $H(\omega)$ at detuning $\omega$ from the optical carrier frequency $\omega_0$---and additive, statistically-stationary, phase-insensitive Gaussian noise $\hat{N}(t)$ with noise spectrum $S_{N} (\omega)=\int d \tau \langle \hat{N}^{\dagger}(t+\tau) \hat{N}(t) \rangle e^{-i \omega \tau}$.  (b) Two unit-cell configurations.   Configuration 1:  20\,dB of attenuation at zero detuning ($\omega = 0$) followed by 20-dB-gain loss-compensating amplification at zero detuning.  Configuration 2: the  20\,dB gain system preceding the 20\,dB attenuation system.   The systems in both configurations include their associated noise sources (not shown).}
	\label{Fig1}
\end{figure}

At power levels below the threshold for significant nonlinear effects, the channel model for fiber-optic communication is a continuous-time LTI filter followed by additive Gaussian noise.  The quantum model from which the fiber channel's HSW capacity can be derived then takes the form shown in Fig.~\ref{Fig1}(a) for quasimonochromatic operation that is subject to an average power constraint.  Here, $\hat{E}_{\rm in}(t)$ and $\hat{E}_{\rm out}(t)$ are baseband $\sqrt{\mbox{photons/sec}}$-units field operators at the channel's input and output, both of which have $\delta$-function commutators:
\begin{equation}
[\hat{E}_{J}(t),\hat{E}_J^\dagger(u)] = \delta(t-u),\mbox{ for $J$ = in, out}.
\label{time-commutators}
\end{equation}
The positive-frequency input and output field operators are thus $\hat{E}_{\rm in}(t)e^{-i\omega_0t}$ and $\hat{E}_{\rm out}(t)e^{-i\omega_0t}$, where $\omega_0$ is the optical carrier frequency of the quasimonochromatic---bandwidth $\Delta \omega \ll \omega_0$---input-field excitation. 
The input-output relation for the Fig.~\ref{Fig1}(a) channel is therefore
\begin{equation}
\label{transformation-eqn}
\hat{E}_{\textrm{out}}(t)=\int\!{\rm d}\tau\,\hat{E}_{\textrm{in}}(\tau) h (t-\tau)+\hat{N}(t),
\end{equation}
where $h(t)$ is the baseband channel's impulse response, which we will assume to be causal ($h(t) = 0$ for $t<0$) and stable ($\int\!{\rm d}t\,|h(t)| < \infty$), and $\hat{N}(t)$ is a baseband noise operator.  The filter's stability ensures that its frequency response,
\begin{equation}
H(\omega) = \int\!{\rm d}t\,h(t)e^{i\omega t},
\end{equation}
exists and provides the frequency-domain version of the input-output relation from Eq.~(\ref{transformation-eqn}): 
\begin{equation}
\hat{\mathcal{E}}_{\rm out}(\omega) = H(\omega)\hat{\mathcal{E}}_{\rm in}(\omega) + \hat{\mathcal{N}}(\omega),
\label{eq:fourier}
\end{equation}
where
\begin{equation}
\hat{\mathcal{E}}_J(\omega) = \int\!{\rm d}t\,\hat{E}_J(t)e^{i\omega t},\mbox{ for $J = \mbox{in, out}$},
\end{equation}
and a similar Fourier transform relates $\hat{\mathcal{N}}(\omega)$ to $\hat{N}(t)$.  The presence of the noise operator is required in order to ensure that the output field operator has the proper $\delta$-function commutator.  In particular, because Eq.~(\ref{time-commutators}) implies
that
\begin{equation}
[\hat{\mathcal{E}}_J(\omega),\hat{\mathcal{E}}^\dagger_J(\omega')] = 2\pi\delta(\omega-\omega'),\mbox{ for $J = \mbox{in, out}$},
\label{freq-commutators}
\end{equation}
we have that
\begin{equation}
[\hat{\mathcal{N}}(\omega),\hat{\mathcal{N}}^\dagger(\omega')] = 
2\pi(1-|H(\omega)|^2)\delta(\omega-\omega').
\label{noise-commutator}
\end{equation}
At frequencies $\omega \in \Omega_{\rm att}$ for which the filter is attenuating ($|H(\omega)| \le 1$), Eqs.~(\ref{eq:fourier}) and (\ref{noise-commutator}) are similar to what we have reported earlier for the single-mode attenuating channel.  Likewise, at frequencies $\omega \in \Omega_{\rm amp}$ for which the filter is amplifying ($|H(\omega)|>1$), these equations are similar to those for the single-mode amplifying channel.  All that remains to complete our channel model is to specify the state associated with the noise operator $\hat{N}(t)$ and to choose some representative frequency responses for our numerical evaluations of the attenuator-amplifier and amplifier-attenuator unit cells in Fig.~\ref{Fig1}(b).  

For our noise models we shall assume that the channel represented by each filter in Fig.~\ref{Fig1}(b) has the minimum possible noise associated with quasimonochromatic operation in thermal equilibrium at temperature $T$\,K, in which case $\hat{N}(t)$ can be taken to be in a zero-mean, statistically-stationary Gaussian state that is completely determined by its phase-insensitive correlation function
\begin{equation}
R_{N}(\tau) = \langle \hat{N}^\dagger(t+\tau)\hat{N}(t)\rangle = \int\!\frac{{\rm d}\omega}{2\pi}\,S_N(\omega)e^{i\omega\tau},
\end{equation}
where
\begin{equation}
S_N(\omega) = \left\{\begin{array}{ll}
\frac{\displaystyle 1-|H(\omega)|^2}{\displaystyle e^{\hbar\omega_0/k_BT} -1}, & \mbox{for $\omega \in \Omega_{\rm att}$} \\[.1in]
\frac{\displaystyle |H(\omega)|^2-1}{1-\displaystyle e^{-\hbar\omega_0/k_BT}}, & \mbox{for $\omega \in \Omega_{\rm amp}$},\end{array}\right.
\label{spectra}
\end{equation}
with $k_B$ being Boltzmann's constant.  Now, to enforce the quasimonochromatic condition---which justifies using $e^{\pm\hbar\omega_0/k_BT}$ in (\ref{spectra}) instead of $e^{\pm\hbar(\omega_0+\omega)/k_BT}$---we shall assume that $H(\omega)$ is narrowband, in comparison with $\omega_0$, such as would be the case for a dense wavelength-division multiplexing (DWDM) filter \cite{Chen,DeCusatis,Berthold}.  In particular, for our numerical work we will employ the fourth-order Butterworth filter, for which $|H(\omega)| = H_0/(1+(\omega/\omega_c)^8)$, where $H_0 \le 1$ is an attenuating filter, $H_0 > 1$ is an amplifying filter, and $\omega_c\ll \omega_0$ enforces the quasimonochromatic condition on the channel filter that will imply a similar quasimonochromatic constraint on $\hat{E}_{\rm in}(t)$'s capacity-achieving excitation spectrum.

\section{Capacities of bosonic LTI channels}
\label{capacities}

To determine the HSW capacity of the LTI channel specified above, we begin by introducing a discretization based on transmitting a stream of $T_s$-sec long continuous-time symbols that are bracketed by $\Delta T_s$-sec long guard bands. In particular, we will assume that the input field $\hat{E}_{\rm in}(t)$ is only in a non-vacuum state when $|t -n(T_s + \Delta T_s)|\le T_s/2$, for integer $n$.  Likewise, after offsetting the receiver's clock by the filter's group delay, we will assume that the receiver only measures the output field  $\hat{E}_{\rm out}(t)$ when $|t -n(T_s + \Delta T_s)|\le T_s/2$, for integer $n$.  By taking $T_s$ to greatly exceed the filter's bandwidth $\omega_c/2\pi$, we can choose a fixed $\Delta T_s$ large enough to ignore intersymbol interference while maintaining $\Delta T_s \ll T_s$ \cite{Gallager}.  It follows that we can focus our attention on a single $n$ value in our discretization.  So, using the $n=0$  operator-valued Fourier series representations,
\begin{equation}
\hat{E}_{\textrm{in}}(t) =\sum_{k} \hat{a}_{k}^{\textrm{in}} \frac{\exp\!\left (-i 2 \pi k t/T_s \right )}{\sqrt{T_s}}, \mbox{ for $|t| \le T_s/2$},
\end{equation}
and 
\begin{equation}
\hat{E}_{\textrm{out}}(t)=\sum_{k} \hat{a}_{k}^{\textrm{out}} \frac{\exp\!\left (-i 2 \pi k t/T_s \right )}{\sqrt{T_s}}, \mbox{ for $|t|\le T_s/2$},
\end{equation}
we obtain
\begin{equation}
\hat{a}_k^{\rm out} = H(2\pi k/T_s)\hat{a}_k^{\rm in} + \hat{n}_k,
\end{equation}
where 
\begin{equation}
\hat{N}(t) =\sum_{k} \hat{n}_{k} \frac{\exp\!\left (-i 2 \pi k t/T_s \right )}{\sqrt{T_s}}, \mbox{ for $|t| \le T_s/2$}.
\end{equation}
The statistics for $\hat{N}(t)$ given in the previous section together with the high time-bandwidth condition $T_s\omega_c/2\pi \gg 1$ imply that the noise operator's Fourier series is also its Karhunen-Lo\`{e}ve series, so that the $\{\hat{n}_k\}$ are in a product state that is Gaussian, zero-mean, and completely characterized by 
\begin{equation}
\langle \hat{n}_k^\dagger\hat{n}_j\rangle = S_N(\omega_k)\delta_{kj},
\end{equation} 
where $\omega_k = 2\pi k/T_s$ and $\delta_{kj}$ is the Kronecker delta function. The discretized capacity problem is then to maximize the Holevo information subject to the average photon-flux constraint \cite{footnote1}
\begin{equation}
\frac{1}{T_s + \Delta T_s}\sum_k\bar{n}(\omega_k) \le P,
\label{fluxconstraint}
\end{equation}
where $\bar{n}(\omega_k) = \langle \hat{a}_k^{{\rm in}\dagger}\hat{a}_k^{\rm in}\rangle$.  The results of \cite{GPCH13,GHG13} imply that the discretized-channel's HSW capacity is achieved by coherent-state encoding.  For such encoding, the Holevo information rate (in bits/sec) is 
\begin{eqnarray}
\lefteqn{\chi(P) = }\nonumber \\[.05in]
&& \!\!\frac{\displaystyle\sum_k \left\{g[|H(\omega_k)|^2\bar{n}(\omega_k) + S_N(\omega_k)] - g[S_N(\omega_k)]\right\}}{T_s+\Delta T_s}\!,
\end{eqnarray}
and the constrained maximization of $\chi(P)$ can be accomplished by a Lagrange multiplier technique, as was done for the multiple-spatial-mode, broadband, pure-loss channel in \cite{ultimate-free-space} and for the beam splitter and amplifier cascade channels in \cite{PMG14}.  Passing to the limit $T_s\rightarrow\infty$ with $\Delta T_s$ fixed then yields the LTI channel's HSW capacity:
\begin{eqnarray}
\lefteqn{C_{\rm HSW}(P) = }\nonumber \\[.05in]
&&\int\!\frac{{\rm d}\omega}{2\pi}\,\left\{[g[|H(\omega)|^2\bar{n}(\omega) + S_N(\omega)] - g[S_N(\omega)]\right\},
\end{eqnarray}
with average photon-number distribution given by 
\begin{eqnarray}
\lefteqn{\bar{n}(\omega) =} \nonumber \\[.05in]
&&\!\!\max\!\left\{\left[(e^{\beta/|H(\omega)|^2}  -1)^{-1} - S_N(\omega)\right]/|H(\omega)|^2,0\right\}\!,
\end{eqnarray}
where the Lagrange multiplier $\beta$ is chosen to saturate the photon-flux bound 
\begin{equation}
\int\!\frac{{\rm d}\omega}{2\pi}\,\bar{n}(\omega) \le P.
\end{equation} 

The homodyne and heterodyne capacities---to which we will compare the preceding HSW capacity---presume coherent-state encoding. 
Hence their capacities are well known, because homodyne and heterodyne measurements convert the Fig.~\ref{Fig1}(a) model into classical LTI channels with additive Gaussian noise.  In particular, assuming unity homodyne and heterodyne efficiencies, the homodyne channel corresponding to Fig.~\ref{Fig1}(a) has an input that is a real-valued, classical, photon-units field $E^{\rm hom}_{\rm in}(t)$ and an output that is a real-valued, classical, photon-units field $E^{\rm hom}_{\rm out}(t)$.  The homodyne channel's input-output relation is then
\begin{equation}
E^{\rm hom}_{\rm out}(t) = \int\!{\rm d}\tau\,E^{\rm hom}_{\rm in}(\tau)h(t-\tau) + N_{\rm hom}(t),
\end{equation}
where $N_{\rm hom}(t)$ is a stationary, zero-mean, real-valued Gaussian random process with spectral density $S_{N_{\rm hom}}(\omega) = (2S_N(\omega) +1)/4$ \cite{footnote2}.   The corresponding channel model for heterodyne detection has complex-valued, classical, photon-units input and output fields that are related by 
\begin{equation}
E^{\rm het}_{\rm out}(t) = \int\!{\rm d}\tau\,E^{\rm het}_{\rm in}(\tau)h(t-\tau) + N_{\rm het}(t),
\end{equation}
where $N_{\rm het}(t)$ is a stationary, zero-mean, isotropic, complex-valued Gaussian random process with spectral density $S_{N_{\rm het}}(\omega) = (S_N(\omega) +1)/2$.  Standard Shannon theory results now lead to the following homodyne and heterodyne capacities \cite{Gallager}
\begin{equation}
C_{\rm hom}(P) = \int\!\frac{{\rm d}\omega}{2\pi}\,\frac{1}{2}\log_2\!\left(1 + \frac{\bar{n}_{\rm hom}(\omega)|H(\omega)|^2}{S_{N_{\rm hom}}(\omega)}\right),
\end{equation}
where 
\begin{equation}
\bar{n}_{\rm hom}(\omega) = \max\left(\beta_{\rm hom}/2 - S_{N_{\rm hom}}(\omega)/|H(\omega)|^2,0\right),
\label{homwater}
\end{equation}
with the Lagrange multiplier $\beta_{\rm hom}$ chosen to give
\begin{equation}
\int\!\frac{{\rm d}\omega}{2\pi}\,\bar{n}_{\rm hom}(\omega) = P,
\label{homflux}
\end{equation}
and
\begin{equation}
C_{\rm het}(P) = \int\!\frac{{\rm d}\omega}{2\pi}\,\log_2\!\left(1 + \frac{\bar{n}_{\rm het}(\omega)|H(\omega)|^2}{2S_{N_{\rm het}}(\omega)}\right),
\end{equation}
where 
\begin{equation}
\bar{n}_{\rm het}(\omega) = \max\left(\beta_{\rm het} -2S_{N_{\rm het}}(\omega)/|H(\omega)|^2,0\right),
\label{hetwater}
\end{equation}
with the Lagrange multiplier $\beta_{\rm het}$ chosen to give
\begin{equation}
\int\!\frac{{\rm d}\omega}{2\pi}\,\bar{n}_{\rm het}(\omega) = P.
\end{equation}
From Eqs.~(\ref{homwater}) and (\ref{hetwater}) it is apparent that the capacity achieving photon-flux spectra for homodyne and heterodyne detection have ``water-filling'' interpretations, e.g., the photon-flux for homodyne detection is allocated across detuning frequencies keeping $\bar{n}(\omega) +  S_{N_{\rm hom}}(\omega)/|H(\omega)|^2$ constant while satisfying Eq.~(\ref{homflux}) \cite{Gallager,Cover}.

\section{Capacities for the unit-cell configurations}
\label{capacities-cell}

Here we will calculate and compare the homodyne, heterodyne, and HSW capacities for the unit-cell configurations shown in Fig.~\ref{Fig1}(b).   We will assume that the amplifying and attenuating components of these configurations have a common normalized frequency response 
\begin{equation}
\bar{H}(\omega) =  \frac{H_{\textrm{amp}} (\omega)}{\max_{\omega} \vert H_{\textrm{amp}} (\omega) \vert  }=\frac{H_{\textrm{atten}} (\omega)}{\max_{\omega} \vert H_{\textrm{atten}} (\omega) \vert },
\end{equation}
given by the fourth-order Butterworth filter that was introduced below (\ref{spectra}).  The amplifier's peak gain, $\max_{\omega}  \vert H_{\textrm{amp}} (\omega) \vert $, which occurs at zero detuning, will be taken to exactly compensate for the attenuator's minimum attenuation, $\max_{\omega}  \vert H_{\textrm{atten}} (\omega) \vert$, which also occurs at that frequency.  Furthermore, as in Sec.~III, both the amplifying and attenuating filters will be taken to have the minimum possible noise associated with quasimonochromatic operation in thermal equilibrium at temperature $T$\,K.  

Finding the unit-cell capacities is actually quite simple, given the results from Sec.~III.  The frequency domain input-output relation for configuration~1 is  
\begin{equation}
\hat{\mathcal{E}}_{\rm out}(\omega) =  H_{\rm amp}(\omega)[H_{\rm att}(\omega)\hat{\mathcal{E}}_{\rm in }(\omega) + \hat{\mathcal{N}}_{\rm att}(\omega)] + \hat{\mathcal{N}}_{\rm amp}(\omega),
\end{equation}
which can be reduced to 
\begin{equation}
\hat{\mathcal{E}}_{\rm out}(\omega) = \bar{H}^2(\omega)\hat{\mathcal{E}}_{\rm in}(\omega) + \hat{\mathcal{N}}_{c1}(\omega),
\end{equation}
where the spectrum associated with the noise operator $\hat{\mathcal{N}}_{c1}(\omega)$ is 
\begin{equation}
S_{N_{c1}}(\omega) = |H_{\rm amp}(\omega)|^2S_{N_{\rm att}}(\omega) + S_{N_{\rm amp}}(\omega),
\end{equation} 
with 
\begin{eqnarray}
\lefteqn{S_{N_{\rm amp}}(\omega) = }\nonumber \\[.05in]
&& \left\{\begin{array}{ll}
\frac{\displaystyle 1-|H_{\rm amp}(\omega)|^2}{\displaystyle e^{\hbar\omega_0/k_BT} -1}, & \mbox{for $|H_{\rm amp}(\omega)| \le 1$} \\[.1in]
\frac{\displaystyle |H_{\rm amp}(\omega)|^2-1}{\displaystyle 1-e^{-\hbar\omega_0/k_BT}}, & \mbox{for $|H_{\rm amp}(\omega)| > 1$},\end{array}\right.
\end{eqnarray}
and
\begin{equation}
S_{N_{\rm att}}(\omega) = 
\frac{\displaystyle 1-|H(\omega)|^2}{\displaystyle e^{\hbar\omega_0/k_BT} -1}. 
\end{equation}
Similarly, the frequency domain input-output relation for configuration~2 can be written as
\begin{equation}
\hat{\mathcal{E}}_{\rm out}(\omega) = \bar{H^{2}}(\omega)\hat{\mathcal{E}}_{\rm in}(\omega) + \hat{\mathcal{N}}_{c2}(\omega),
\end{equation}
where the spectrum associated with the noise operator $\hat{\mathcal{N}}_{c2}(\omega)$ is 
\begin{equation}
S_{N_{c2}}(\omega) = |H_{\rm att}(\omega)|^2S_{N_{\rm amp}}(\omega) + S_{N_{\rm att}}(\omega).
\end{equation} 

The preceding results demonstrate that unit-cell configurations~1 and 2 are both attenuating channels, in the sense of Fig.~\ref{Fig1}(a), but they are \emph{not} minimum-noise attenuating channels.  Furthermore, both configurations have the same frequency response for their signal transmission, but, because configuration~1 has a higher noise spectrum, configuration~2's HSW, homodyne, and heterodyne capacities will exceed their configuration~1 counterparts.  

\begin{figure}[htb!]
\centering
	\includegraphics[width=\columnwidth]{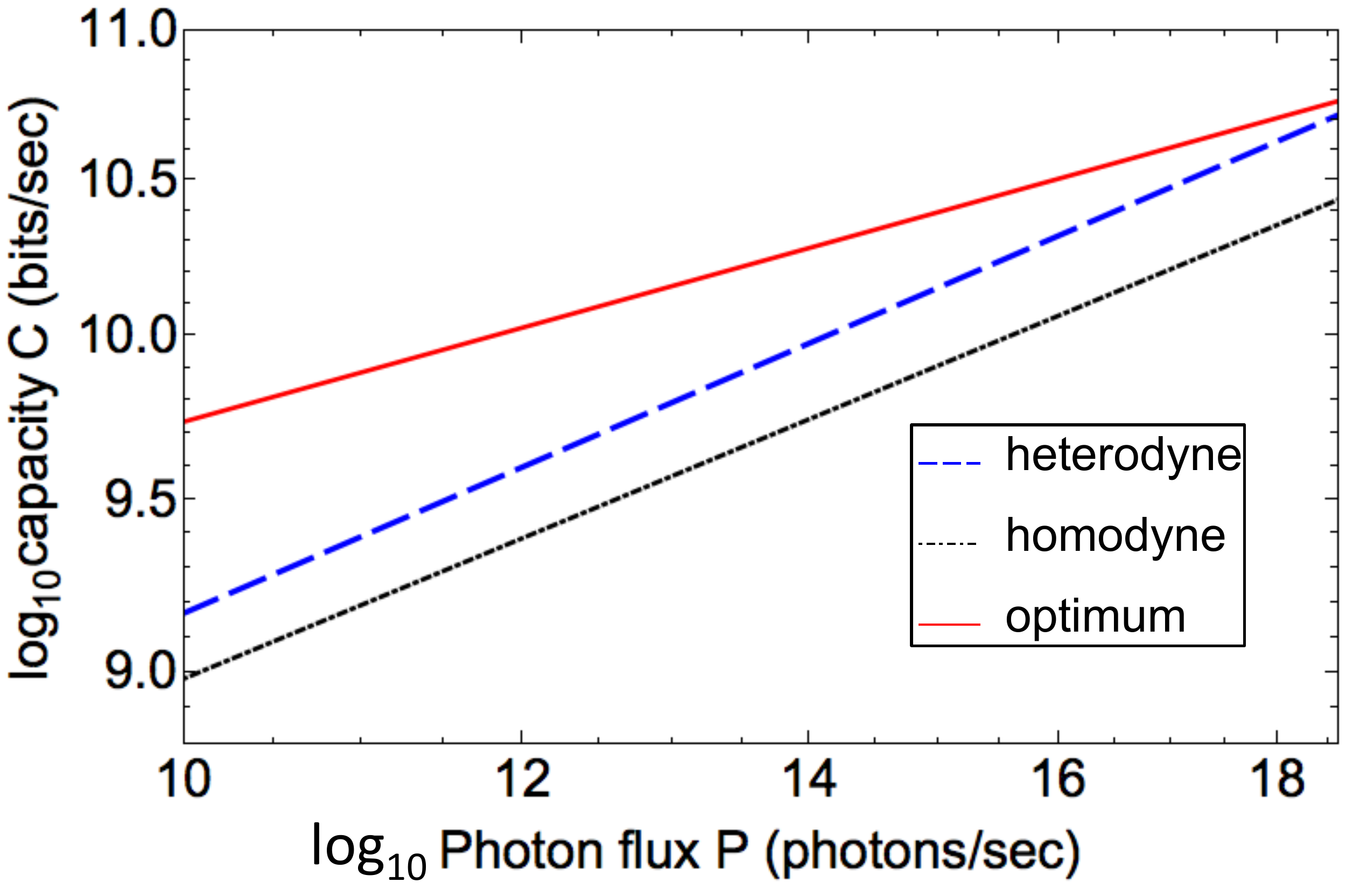}
	\caption{(color online). Heterodyne, homodyne, and HSW capacities versus transmitted photon flux for configuration~1.   The plots assume  1550\,nm center wavelength and operation at $T=300\,$K with $\omega_c/2\pi = 20\,$GHz fourth-order Butterworth filters.}
	\label{Fig2}
\end{figure}

\begin{figure}[htb!]
\centering
	\includegraphics[width=\columnwidth]{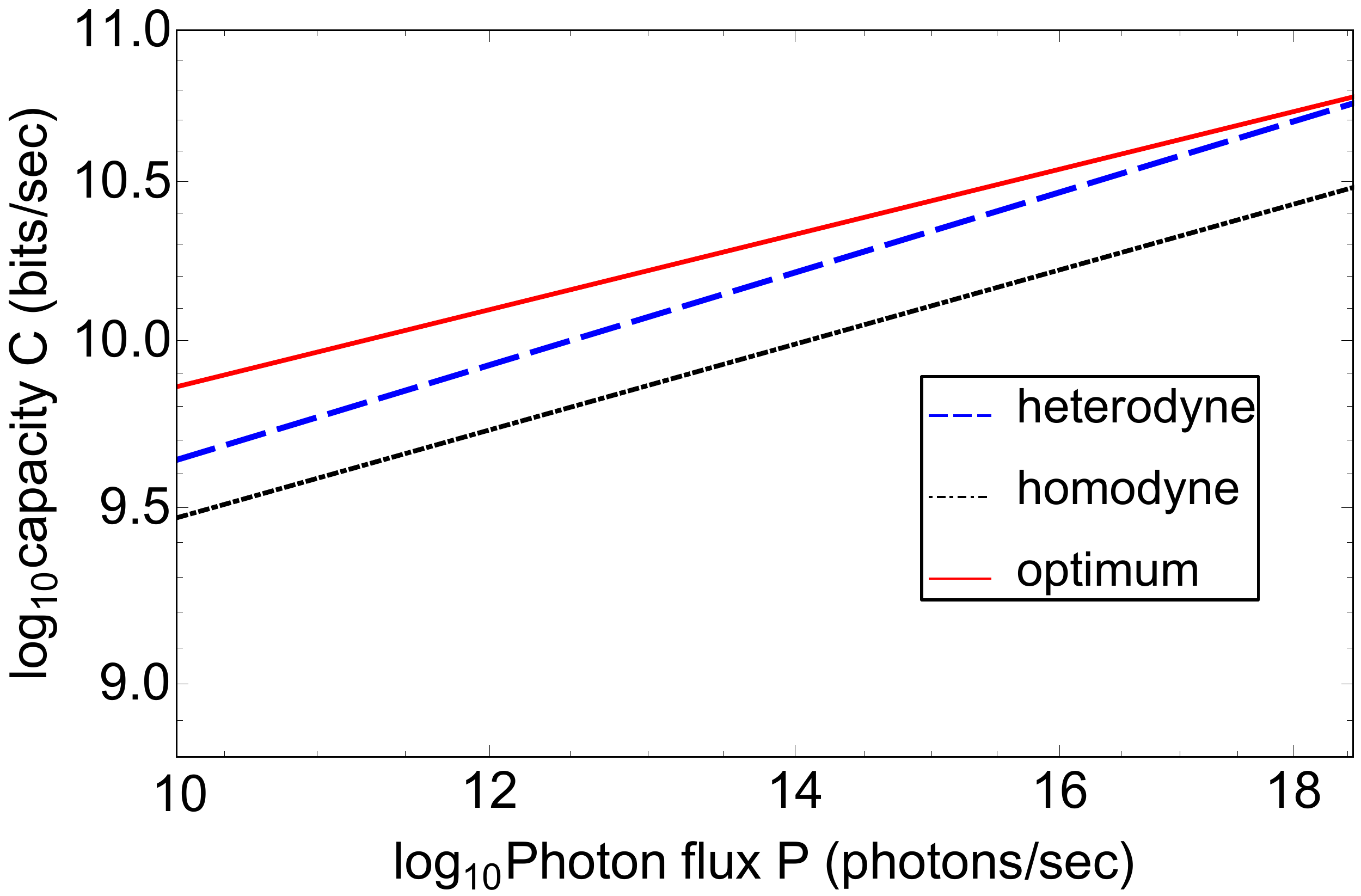}
	\caption{(color online). Heterodyne, homodyne, and HSW capacities versus transmitted photon flux for configuration~2.   The plots assume  1550\,nm center wavelength and operation at $T=300\,$K with $\omega_c/2\pi = 20\,$GHz fourth-order Butterworth filters.}
	\label{Fig3}
\end{figure}

In Fig.~\ref{Fig2}, we plot the HSW, heterodyne, and homodyne capacities versus the transmitted photon flux for configuration~1, where we have assumed a 1550\,nm center wavelength and operation at $T=300\,$K with $\omega_c/2\pi = 20\,$GHz fourth-order Butterworth filters.; Fig.~\ref{Fig3} contains the corresponding capacity plots for configuration~2.  As expected, configuration~2's capacities exceed their configuration~1 counterparts because of the latter's noise spectra being higher than the former's.  For both configurations, the flat passband and steep-skirted behavior of the Butterworth filter makes the heterodyne capacity approach the HSW capacity at high photon-flux levels, while the heterodyne capacity exceeds the homodyne capacity for all photon fluxes shown in the figures. \\

\section{Conclusions}
\label{conclusion}

In this work, we have presented a quantum mechanical model for optical communication through LTI bosonic channels with additive Gaussian noise, and we have reported a framework for evaluating their HSW capacities. Such bosonic channels can represent the effects of quantum amplification or attenuation in a thermal-noise environment, as encountered in fiber-optic communication at power levels below the threshold for significant nonlinear effects.  Our numerical work provides a comparison between the optimum-reception capacity, for a representative candidate filter, with corresponding results for heterodyne and homodyne detection over the same channel.  Although carried out for single-wavelength operation with fiber propagation in mind, our results can easily be extended to the multi-wavelength case of DWDM transmission~\cite{Chen,DeCusatis, Kahn, Patel}.  Likewise, our single unit-cell evaluations can easily be extended to treat a chain of such unit cells.  With additional work, to account for fading, our bosonic channel model could be applied to free-space optical communication with thermal noise~\cite{Niv1,Niv2}.  

\section{Acknowledgements}

This research was supported by AFOSR grant number FA9550-14-1-0052.


\begin{thebibliography}{99}
\bibitem{footnote1} Because $\hat{E}_{\rm in}(t)$'s excitation is quasimonochromatic, this constraint limits its average power to at most $\hbar\omega_0P$.
\bibitem{footnote2} Here, without appreciable loss of generality, we have assumed that the impulse response $h(t)$ is real valued.
\end{thebibliography}
\end{document}